\documentclass[final,5p,times,twocolumn,square]{elsarticle}
\usepackage[utf8]{inputenc}
\usepackage{verbatim}
\usepackage{amsmath}
\usepackage{mathtools}
\usepackage{gensymb}
\usepackage{amssymb}
\usepackage{soul}
\usepackage{xcolor}
\usepackage[normalem]{ulem}

\usepackage{multirow}

\usepackage{soul}

\journal{Ultramicroscopy}

\begin{document}
\begin{frontmatter}

\title{Towards reliable electrical measurements of superconducting devices inside a transmission electron microscope}

\author[label1]{Joachim Dahl Thomsen\corref{cor1}}
\cortext[cor1]{j.thomsen@fz-juelich.de}

\author[label1]{Michael I. Faley}

\author[label1]{Joseph Vimal Vas}

\author[label1]{Alexander Clausen}
            
\author[label1]{Thibaud Denneulin}

\author[label2]{Dominik Biscette}

\author[label2]{Denys Sutter}
       
\author[label1]{Peng-Han Lu\corref{cor2}}
\cortext[cor2]{p.lu@fz-juelich.de}
           
\author[label1]{Rafal E. Dunin-Borkowski}

\affiliation[label1]{organization={Ernst Ruska-Centre for Microscopy and Spectroscopy with Electrons, Forschungszentrum Jülich}, 
            city={52428 Jülich},
            country={Germany}}

\affiliation[label2]{organization={condenZero AG}, 
            city={8050 Zürich},
            country={Switzerland}}
            
\begin{abstract}
Correlating structure with electronic functionality is central to the engineering of quantum materials and devices whose properties depend sensitively on disorder. Transmission electron microscopy (TEM) offers high spatial resolution together with access to structural, electronic, and magnetic degrees of freedom. However, \textit{operando} electrical transport measurements on functional quantum devices remain rare, particularly at liquid helium temperature. Here, we demonstrate electrical transport measurements of niobium nitride (NbN) devices inside a TEM using a continuous-flow liquid-helium-cooled sample holder. By optimizing a thermal radiation shield to limit radiation from the nearby pole pieces of the objective lens, we achieve an estimated base sample temperature of 8–9~K, as inferred from the superconducting transition temperatures of our devices. We find that both electron beam illumination and objective lens excitation perturb the superconducting state. In addition, we evaluate the imaging capabilities and stability of the sample holder at low temperature by imaging the magnetic domain structure of the van der Waals ferromagnet CrBr$_3$. Finally, we perform calculations that underscore the importance of cryo-shielding for minimizing thermal radiation onto the device. This capability enables correlative low-temperature TEM studies, in which structural, spectroscopic, and electrical transport data can be obtained from the same device, thereby providing a platform for probing the microscopic origins of quantum phenomena.
\end{abstract}

\begin{highlights}
\item Superconducting device measurements inside a transmission electron microscope.
\item Enabled using a continuous-flow liquid-helium-cooled sample holder.
\item Efficient thermal radiation shielding is essential to achieve a low specimen temperature.
\item Electron-beam-induced heating perturbs the superconducting state. 
\end{highlights}

\begin{keyword}
Transmission electron microscopy \sep Superconductivity \sep Cryo-TEM \sep \textit{Operando} experiments
\end{keyword}

\end{frontmatter}

\section{Introduction}

Understanding the interplay between atomic-scale structure and quantum phenomena is a central challenge in condensed matter physics \cite{ko2023understanding}. In quantum materials, strain or defects can significantly influence macroscopic properties such as magnetism, topology, and superconductivity \cite{tokura2017emergent, mazza2024embracing}, motivating high-resolution structural characterization that is directly correlated with functional property measurements for the design, synthesis, and understanding of quantum materials and devices \cite{moler2017imaging, minor2019cryogenic, bianco2021atomic, hart2021seeing}.

Transmission electron microscopy (TEM) offers unmatched spatial resolution for probing such structural features, while enabling the mapping of local electric and magnetic fields \textit{via} techniques such as Lorentz TEM, off-axis electron holography, and four-dimensional scanning TEM (4D-STEM) \cite{midgley2009electron, han2025electric, chen2022lorentz, kang2025large}. In addition, electron energy-loss spectroscopy (EELS) provides local information about orbital occupancy \cite{cantoni2014orbital}, charge transfer \cite{zhao2018direct}, and spin and orbital angular moments \cite{ali2025visualizing}. Together, these capabilities enable probing of the lattice, spin, charge, and orbital degrees of freedom in quantum materials using a single instrument.

\begin{figure*}[t]
\centering
\includegraphics[width=1\linewidth]{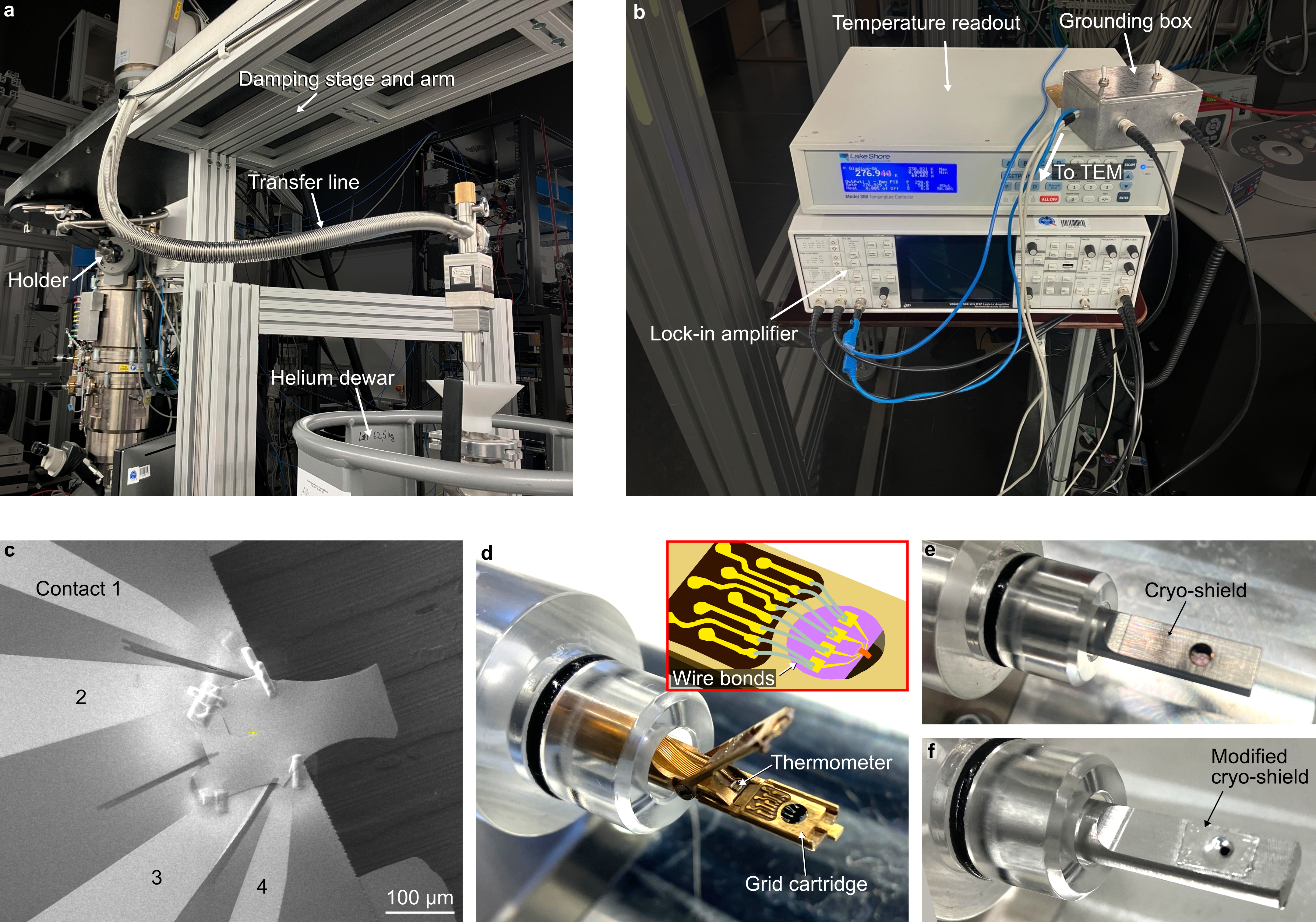}
\caption{\label{fig:equipment}
		\textbf{Measurement setup overview.} 
		\textbf{(a)} Photograph of the helium dewar, helium transfer line inserted into the back of the TEM holder, and the arm of the damping stage that supports the helium transfer line.
        \textbf{(b)} Photograph of the temperature sensor readout, lock-in amplifier, and grounding box. A diagram of the electrical measurement setup is given in Fig.~S2. In (a, b) the background has been blacked out to emphasize the equipment used in this work. 
        \textbf{(c)} SEM image of NbN device 1. The numbering of the electrodes is indicated on the image. 
        \textbf{(d)} TEM sample holder with opened tip. The grid cartridge with the TEM grid has been mounted onto the holder. The location of the thermometer is indicated. The TEM sample holder is placed on a holder stand. The inset shows a diagram of the TEM grid, the grid cartridge, and the wire bonds connecting both together.
        \textbf{(e, f)} Photographs of the TEM sample holder with the tip closed and encased in the (e) regular and (f) modified cryo-shield. 
		}
\end{figure*}

Visualizing quantum properties often requires ultra-low sample temperatures. Liquid-helium-cooled TEM imaging has, for example, been used to visualize magnetization textures \cite{han2019topological} and superconducting flux vortices \cite{harada1992real, harada1996direct}. Recent technological developments are enabling atomic-resolution imaging at temperatures around 30~K \cite{mun2024atomic, rennich2025ultracold}. However, \textit{operando} electronic transport measurements of quantum devices remain rare. Such measurements could, for instance, enable direct correlation between microstructure and superconducting behavior, and allow unambiguous identification of superconductivity without relying on vortex imaging, which requires restrictive experimental conditions. Moreover, the difference between the actual sample temperature and the temperature recorded by the sample holder can be substantial \cite{kumar2024calibrating}. Precise temperature calibration is therefore essential for accurate \textit{operando} measurements of quantum devices.

Here, we demonstrate electrical transport measurements on niobium nitride (NbN) devices performed inside a TEM using a continuous-flow liquid-helium-cooled sample holder, which supports nanometer-scale imaging at temperatures close to liquid helium temperature \cite{kim2025ultralow}. NbN is a type-II superconductor with a superconducting transition temperature ($T_{\mathrm{c}}$) in the range 11–17~K, depending on the sample geometry and synthesis conditions \cite{mathur1972lower, hwang2015transition, hazra2016superconducting, volkov2019superconducting, guo2020fabrication}. Its relatively high $T_{\mathrm{c}}$ makes NbN an ideal material for benchmarking superconducting device measurements in the TEM and calibrating the true sample temperature. We perform such measurements, resulting in an estimated sample base temperature of 8-9~K. Our results demonstrate the importance of efficient thermal shielding for reaching low specimen temperatures, and quantitatively assess how electron-beam illumination and objective-lens excitation perturb electrical biasing experiments. In addition, we evaluate imaging capabilities and stability of the sample holder through videos of Lorentz TEM images of CrBr$_3$, a van der Waals-layered ferromagnet with a Curie temperature of $\sim$37 K \cite{grebenchuk2024topological}. 

Our results show that superconductivity in electrically contacted devices can be realized and detected in the TEM. This capability paves the way for future correlative low-temperature TEM studies in which structural, spectroscopic, and transport information can be acquired from the same specimen.

\section{Experimental details}

\subsection{Helium cooling}

Helium cooling is performed using a continuous-flow liquid-helium-cooled sample holder (condenZero AG, Switzerland), which allows temperature control between a thermometer reading of 4.5~K and room temperature. Liquid helium is transferred from a helium dewar \textit{via} a helium transfer line supported by an arm, as shown in Fig.~\ref{fig:equipment}(a). The helium dewar and support arm are mounted on a vibration damping stage. The over-pressure in the helium dewar, which is typically set to 200-400~mbar, controls the flow of liquid helium through the sample holder. The cooling time from room temperature is {$\sim$}10~min, as shown in Fig.~S1. We estimate that the holding time at low temperature can exceed 24~hours, as it only limited by the amount of helium in the dewar. 

\subsection{Electrical property measurements}

The TEM sample holder has 8 electrical feedthroughs, four of which were connected \textit{via} a breakout box for four-probe resistance measurements. A circuit diagram of the setup is shown in Fig.~S2. A lock-in amplifier was used for resistance measurements to improve the measurement sensitivity (Fig.~\ref{fig:equipment}(b)). An excitation current of 100~nA–1~$\mu$A was applied through the outer contacts (1 and 4; see Fig.~\ref{fig:equipment}(c))
while measuring the voltage across the device (contacts 2 and 3). 
Additional switches and shunt resistors were added to the grounding box (Fig.~\ref{fig:equipment}(b)) to protect the device from spurious electrostatic discharges. Electrical connections between the TEM grid and the grid cartridge are made through wire bonding (Fig.~\ref{fig:equipment}(d) and inset). With 8 electrical feedthroughs available, the connections to each of the four contacts on the TEM grid was doubled in case any of the wire bonds broke during the experiments. Spring contacts on the top lid of the sample holder then connect to the contact pads on the grid cartridge. An inbuilt Cernox thermometer (Lake Shore Cryotronics, USA), shown in Fig.~\ref{fig:equipment}(d), was used to measure the temperature of the holder. The temperature and resistance were logged automatically using custom Python scripts.

Electrical properties were also evaluated in a dedicated cryostat for electrical transport measurements with calibrated temperature control--a Physical Property Measurement System (PPMS, DynaCool, Quantum Design, USA) using a lock-in technique with an excitation current of 10~µA. In the PPMS, the sample chamber wall is cooled, minimizing thermal radiation exchange between the device and chamber walls. For these measurements, the grid cartridges were wire-bonded to PPMS-compatible chips. The wire bonds between the TEM grid and the grid cartridge were identical for the PPMS and TEM measurements. 

\subsection{NbN sample preparation}
Deposition of a 6~$\mu$m-thick NbN film was performed in a home-made pulsed DC magnetron sputtering system from a 50~mm diameter target of 99.5\% pure NbN onto a single-crystalline MgO (001) substrate heated to approximately 650\,$^\circ$C (heater temperature 800\,$^\circ$C). A gas mixture of 99.9999\% pure Ar (90\%) and N$_2$ (10\%) at a total pressure of 0.01~mbar was used during deposition. The NbN films grow epitaxially with the NbN crystal \textit{c}-axis perpendicular to the MgO substrate and exhibit a typical $T_{\mathrm{c}}$ of {$\sim$}16.5~K. After cooling to room temperature, the film was under compressive stress due to the difference in the thermal expansion coefficients of NbN and MgO (4.2$\times$10$^{-6}$ and 1.0$\times$10$^{-5}$~K$^{-1}$, respectively). Under the influence of this stress and due to the relatively low cleavage energy of MgO ({$\sim$}1.2~J/m$^2$ \cite{boffi1976cleavage}), the NbN film fragmented into several flakes of different sizes. Flakes measuring approximately 70~$\times$~150~$\mu$m$^{2}$ were selected and transferred to TEM grids using the nanomanipulator in a Thermo Fisher Scientific Helios Hydra dual beam plasma-focused ion beam (FIB), as shown in Fig.~S3. The flakes were connected to Au current leads on SiN lift-out TEM grids (SiMPore, USA - product no.\ SN100-LFT-AU) using focused-electron-beam-induced W deposition in the plasma-FIB. The $T_{\mathrm{c}}$ of the NbN flakes after mounting to TEM grids is lower than that of NbN films and is typically around 12~K. Two NbN devices were made in total. Due to a degradation of the superconducting properties in thin FIB-prepared NbN lamellae, NbN device 1 was only thinned down using the plasma-FIB to a thickness of $\sim$1~$\mu$m while NbN device 2 was kept at its as-grown 6~$\mu$m thickness. In both cases, the devices were opaque to the electron beam. We have found that films with thicknesses around 100~nm, that are more compatible with TEM imaging, do not lead to fragmentation and delamination from the growth substrate and therefore require alternative growth or transfer methods to fabricate a TEM specimen.

\subsection{CrBr$_3$ sample preparation}
Few-layer graphene (NaturGrafit GmbH, Germany) and CrBr$_3$ flakes (HQ Graphene, Netherlands) were prepared by mechanical exfoliation of bulk crystals onto Si substrates with 90~nm-thick SiO$_2$ using 3M Magic Scotch tape. To protect against oxidation, the CrBr$_3$ flake was encapsulated between few-layer graphene flakes using a dry-transfer method based on a polycarbonate (PC)-coated polydimethylsiloxane (PDMS) stamp mounted on a glass slide. Exfoliation and heterostructure assembly were carried out in a glovebox under an inert N$_2$ atmosphere. The residual PC film was dissolved in chloroform outside the glovebox, and the heterostructure was then transferred to a Si/SiN TEM grid (Norcada Inc., Canada) using cellulose acetate butyrate (CAB) as a polymer handle \cite{schneider2010wedging}. Finally, the CAB film was dissolved in acetone, and the TEM grids were rinsed in isopropanol and dried in air.

\subsection{Transmission electron microscopy}

TEM experiments were performed on an FEI Titan G2 60–300 operated at 300~kV in Lorentz TEM mode, with a residual magnetic field of {$\sim$}2~mT at the sample position. The magnetic field as a function of objective lens excitation was calibrated using a TEM holder equipped with a Hall probe. The microscope hosts a C-TWIN objective lens with a pole piece gap of 11~mm \cite{boothroyd2016fei}. The electron beam current density at different spot sizes was determined from the recorded counts on a direct-detection Gatan K2 camera in counting mode. At low beam currents the measured counts on the camera scale linearly with the actual current density \cite{li2013electron}. For the two smallest spot sizes used in the experiments, we estimated that the beam current approximately doubles each time the spot size is decreased. 

\begin{figure*}[]
    \centering
	\includegraphics[width=0.90\linewidth]{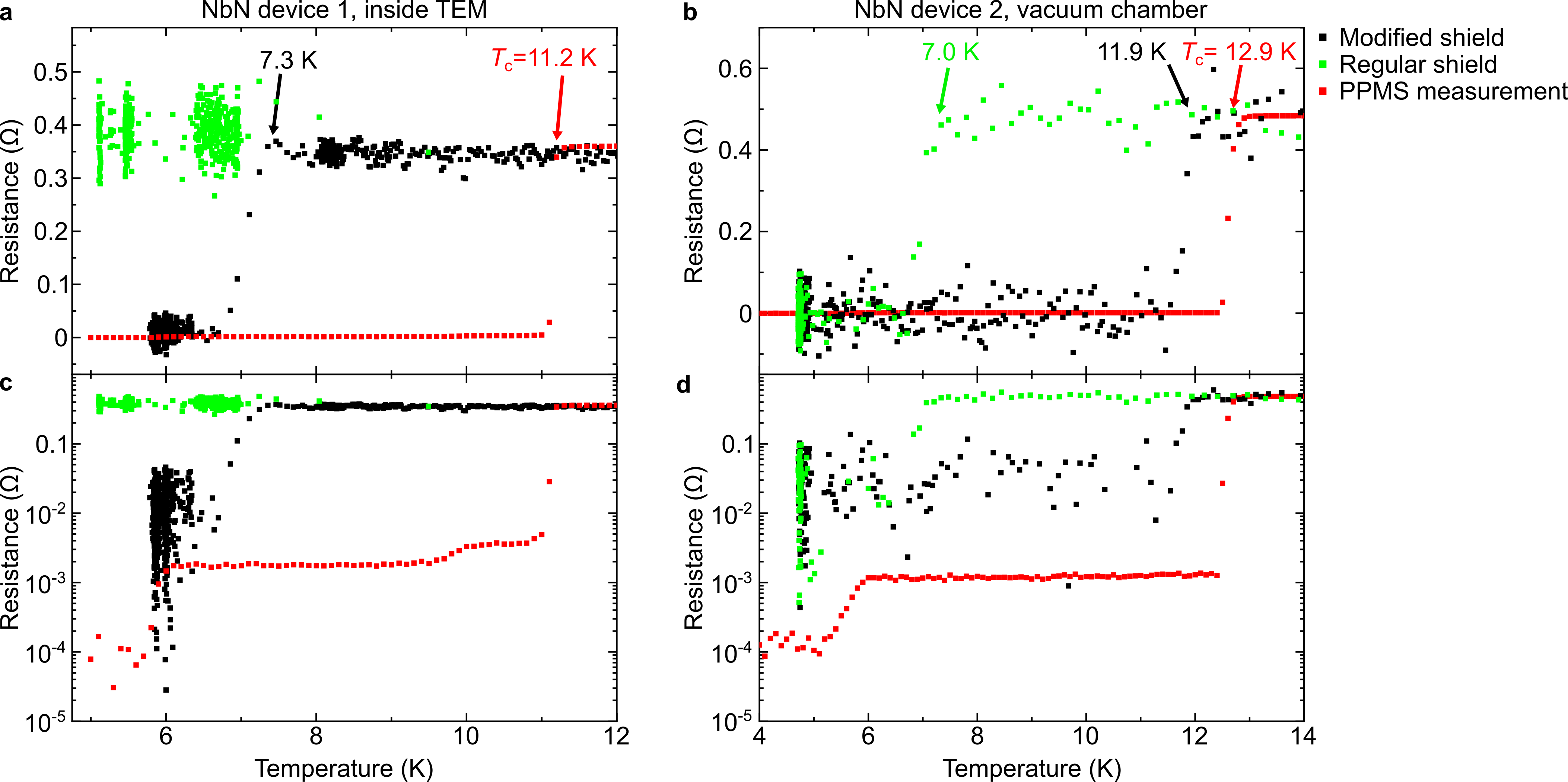}
	\caption{\label{fig:cooling}
		\textbf{Resistance measurements of superconducting NbN devices.} 
        \textbf{(a, c)} Resistance plotted against thermometer temperature for NbN device 1. The measurements were performed in the TEM in LTEM mode without electron beam illumination. The black and green data were obtained using the modified and regular cryo-shield, and using test currents of 500 and 100~nA, respectively. The black and green data were measured during warming and cooling, respectively. The rate of warming can be controlled much more precisely using small over-pressures (0-100~mbar) in the helium dewar. The small number of green data between 7-12~K is due to the fast cooling rate in this temperature range. The fast cooling rate does not affect the presence of a superconducting transition, but would result in an inaccurate measurement of $T_{\mathrm{c}}$ since the thermometer and specimen are not in thermal equilibrium. 
        \textbf{(b, d)} Resistance plotted against thermometer temperature for NbN device~2. The measurements were performed in a vacuum chamber. The black and green data were obtained with the modified and regular cryo-shield, respectively, using a test current of 100~nA. In this case, the cryo-shield was modified by covering the imaging aperture of the cryo-shield entirely with aluminium. The red data in (a-d) were obtained from separate PPMS measurements. 
		}
\end{figure*}

\section{Results and discussion}

\subsection{Thermal radiation and cryo-shielding}
We start by investigating the effect of cryo-shielding and thermal radiation on the lowest achievable specimen temperature without any simultaneous electron beam illumination. Figure~\ref{fig:equipment}(e) shows the as-supplied (“regular”) cryo-shield, which has a 3-mm-diameter opening for TEM imaging. We fabricated a modified cryo-shield by covering both the top and bottom opening with aluminium tape and making a $\sim$0.5~mm diameter hole to reduce the direct line-of-sight to thermal radiation sources, such as the nearby objective lens of the TEM, as shown in Fig.~\ref{fig:equipment}(f). The cryo-shield is cooled by helium flow returning from the holder tip and is expected to be warmer than the thermometer reading. 

Figure~\ref{fig:cooling}(a) shows resistance–temperature measurements for NbN device 1 acquired inside the TEM in LTEM mode (objective lens turned off) with the electron beam blanked. The black data were obtained using the modified cryo-shield and exhibit an apparent superconducting transition at $T_{\mathrm{c}}^{\ast}$=7.3~K. Independent measurements in a PPMS prior to the TEM experiments gave an intrinsic $T_{\mathrm{c}}$=11.2~K (red data). This discrepancy suggests that, under these conditions, the actual specimen temperature is approximately 4~K higher than the thermometer temperature. During this measurement, the thermometer reached a base reading of 5.8~K; applying the same offset yields an estimated specimen temperature of $\sim$10K. The lowest thermometer reading observed across all experiments was 4.5~K. Taken together, these observations suggest that the lowest achievable specimen temperature is 8–9~K when using the modified cryo-shield. We summarize our temperature estimates in Table~1, together with additional data discussed below.

\begin{table*}[]
\centering
\begin{tabular}{p{2.5 cm} p{3.2 cm} p{3.2 cm} p{3.2 cm} p{3.2 cm} c} 
 \hline
  \multirow{2}{*}{\textbf{\,\,\,\,Cryo-shield}} & \multicolumn{2}{c}{\textbf{Inside TEM}} & \multicolumn{2}{c}{\textbf{Inside vacuum chamber}} &\\  
 \cline{2-6}
 & \centering Apparent $T_{\mathrm{c}}^{\ast}$ & \centering Estimated specimen $T$ & \centering Apparent $T_{\mathrm{c}}^{\ast}$ & \centering Estimated specimen $T$ & \\  \hline
 \centering Regular & \centering N/A & \centering $>$11.2 K & \centering 7.0 K & \centering $\sim$11 K & \\ 
 \centering Modified & \centering 7.3 K & \centering 8-9 K & \centering 11.9 K\textsuperscript{\textdagger} & \centering $\sim$6 K\textsuperscript{\textdagger} &\\
 \hline
\end{tabular}
\caption{Measured apparent $T_{\mathrm{c}}^{\ast}$ and estimated specimen temperature in the TEM (beam blanked) and in the vacuum chamber. The specimen temperature was estimated by considering the difference in apparent $T_{\mathrm{c}}^{\ast}$ and the actual $T_{\mathrm{c}}$ measured with the PPMS, and adding this difference to the base thermometer temperature for that measurement. \textsuperscript{\textdagger}The modified heat shield in these cases has a fully covered imaging aperture.}
\label{table:1}
\end{table*}

The green data in Fig.~\ref{fig:cooling}(a) were obtained using the regular cryo-shield. In this case, no superconducting transition was measured, suggesting that when using the regular cryo-shield, thermal radiation keeps the specimen temperature above 11.2~K, even though the lowest thermometer reading was 4.5~K and the electron beam was blanked. The higher noise in the green data arises from the lower measurement current used (see figure caption for details). We systematically varied the measurement current in the range 100~nA–1~µA and did not observe any noticeable change in apparent $T_{\mathrm{c}}$, indicating that Joule heating is negligible under our measurement conditions (see Fig.~S4).

Next, we investigate the impact of the local environment of the sample holder on the specimen temperature by  performing electrical property measurements in a separate vacuum chamber. A photograph of the vacuum chamber is shown in Fig.~S5. 

Figure~\ref{fig:cooling}(b) shows resistance–temperature measurements for NbN device 2 obtained in the vacuum chamber. Using the regular cryo-shield (green data), we detect a superconducting transition with an apparent $T_{\mathrm{c}}^{\ast}$=7~K. Comparison with the PPMS measurement, which shows $T_{\mathrm{c}}$=12.9~K, indicates that the specimen temperature is {$\sim$}6~K higher than the thermometer reading. This value can be improved significantly by modifying the cryo-shield, in this case by entirely covering the imaging aperture with aluminium (black data). The specimen temperature is now only {$\sim$}1~K lower than the thermometer reading, suggesting a specimen temperature of {$\sim$}6~K at the base temperature. This measurement corresponds to a best-case scenario with minimised thermal radiation (no objective lens) and optimal heat shielding (fully covered imaging apertures).

Figure~2(c,d) replots the resistance data from Fig.~2(a,b) on a logarithmic scale. The PPMS measurements show that NbN devices 1 and 2 exhibit a small residual resistance below $T_{\mathrm{c}}$ that vanishes for $T$<5-6~K and lies below the noise floor of the TEM measurements. This residual resistance is approximately 0.5\% (device 1) and 0.2\% (device 2) of the resistance just above $T_{\mathrm{c}}$ and does not affect the temperature estimates reported here. The finite resistance below $T_{\mathrm{c}}$ may arise from parasitic amorphous W deposited during FIB fabrication of the W contacts between the NbN and Au electrodes, thereby forming a partial shunt between contacts. Amorphous W films deposited by FIB can exhibit a $T_{\mathrm{c}}$ of 5-6~K \cite{sadki2004focused, li2008tunability}, consistent with our observations. In future work, weld-free mounting of lamellae could help eliminate such parasitic conduction paths \cite{recalde2024weld}. We also note that the logarithmic plots in Fig.~2(c,d) do not display the negative values observed in the noisy TEM data below $T_{\mathrm{c}}$. As a result, the mean resistance below $T_{\mathrm{c}}$ appears larger than its true value; nevertheless, these plots remain useful for visualizing any residual resistance.

To further probe the superconducting properties of our devices, Fig.~S6 shows the resistance of NbN device 2 as a function of applied magnetic field at $T$=10~K, measured in the PPMS. The device exhibits behavior consistent with superconductivity, with a sharp transition to the normal-state resistance at a critical field of $\pm$2.8~T. Such a high critical resistance suggests that the device could also be used in normal TEM mode with the objective lens fully excited. Together, the resistance measurements as a function of temperature and magnetic field provide strong evidence for superconductivity, with only a very small residual parasitic shunt contribution likely arising from FIB-deposited W.

The difference in specimen temperature between experiments in the TEM and in the vacuum chamber may arise from differences in mechanical coupling between the sample holder and its surrounding hardware. In the TEM, a weak mechanical contact between the cryo-shield and the goniometer can introduce an additional conductive heat path, leading to cryo-shield heating and thereby increasing the radiative load on the specimen. In the vacuum chamber, there is no additional mechanical support beyond the contact between the sample holder O-ring and the chamber entry port. Differences in the radiative environment may also play a role: in the TEM, the sample sits within a narrow pole-piece gap, whereas in the vacuum chamber the specimen is located {$\sim$}10~cm from the chamber walls. This geometry could reduce the view factor to warm surfaces, potentially lowering the thermal radiation reaching the specimen.

These results demonstrate that efficient heat shielding is crucial for achieving a low specimen temperature. In particular, thermal radiation raises the specimen temperature relative to the thermometer reading. The thermometer is located {$\sim$}1~cm away from the specimen position, closer to the helium line, and is therefore cooled more efficiently. A drawback of the modified cryo-shield is that the specimen cannot be tilted significantly before the shield blocks the electron beam. Additional heat shielding using, for example, cryo-blades within the objective lens pole pieces or a cryo-cage surrounding the specimen should be further investigated. Ultimately, in order to achieve the lowest possible specimen temperature, concepts such as liquid-helium-cooled superconducting objective lenses may be necessary \cite{lefranc1982superconducting, weierstall1996electron}.

The accuracy of the $T_{\mathrm{c}}$-measurement could be further improved by more precise temperature control. Ideally, the heating or cooling rates should be sufficiently low that the specimen and thermometer remain in thermal equilibrium. In our measurements, we determine $T_{\mathrm{c}}$ during heating, because in the range 5–20~K the heating rate can be reduced to below 1~K/s by slowly decreasing the over-pressure in the helium dewar. The cooling rate, in contrast, cannot be controlled to the same extent and may exceed 20~K/s depending on the over-pressure \cite{kim2025ultralow}. Typical heating and cooling rates during our experiments are shown in Fig.~S1. 

A possible route to improved temperature control would be to cool the holder to its base temperature and then locally heat the specimen using a micro-electromechanical (MEMS) heater, as can be done during liquid nitrogen-cooling \cite{goodge2020atomic, hart2023operando}. The heater resistance could then serve as a thermometer located at the specimen position. However, MEMS heaters are typically fabricated from Pt, Mo, or other metallic compounds, which do not show a strong temperature variation below 50~K. Hence, there is a need for new MEMS heaters that are capable of reliable operation down to 5~K.

\begin{figure*}[]
	\includegraphics[width=1\linewidth]{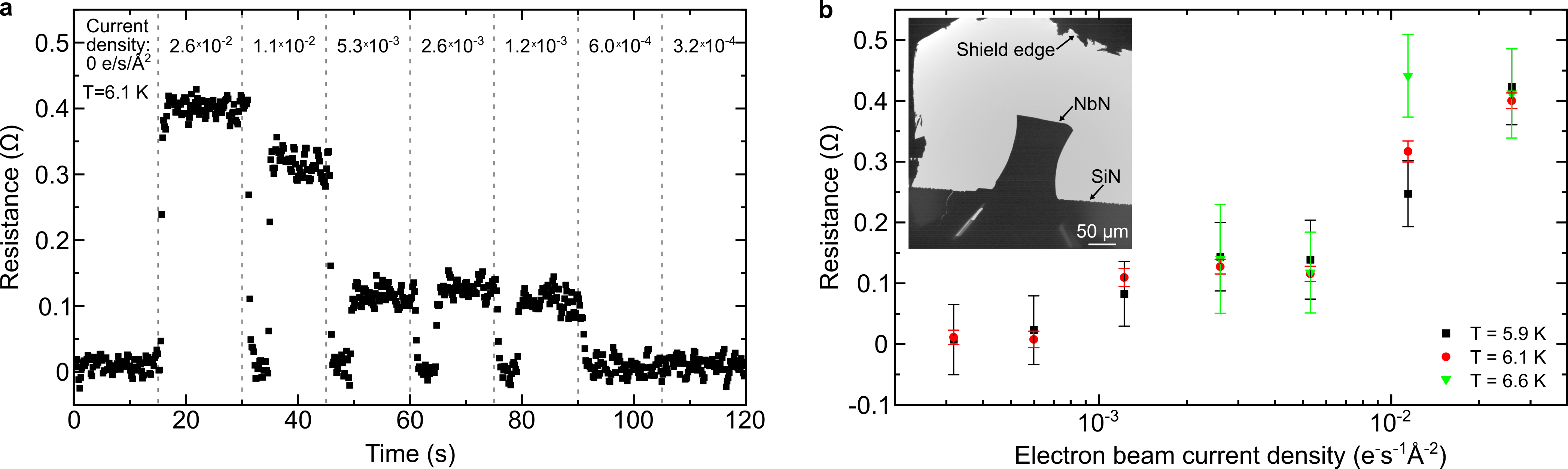}
	\caption{\label{fig:beam}
		\textbf{Effect of TEM imaging.} 
		\textbf{(a)} Resistance plotted against time for NbN device 1 during an experiment where the beam is initially blanked for 15~s, and then the beam current is decreased every 15~s by increasing the spot size. The device becomes superconducting between each spot size change because the electron beam is briefly blanked during the change. The electron beam current density is indicated on the figure. The experiment was performed at a thermometer temperature of 6.1~K.
  		\textbf{(b)} Resistance plotted against electron beam current density for four different experiments performed at slightly different thermometer temperatures between 5.9 and 6.6~K. During each experiment, the temperature was stable with a standard deviation $\sim$0.01~K for on the thermometer readings. The inset shows a TEM image of the device.  
    }
\end{figure*}

\subsection{Electron beam effects}

We next examine the influence of TEM imaging on the electrical properties of the NbN devices. In these experiments, we measure the resistance of NbN device 1 using the modified cryo-shield in Lorentz TEM mode (objective lens turned off). The temperature is held constant during each of the experiments presented in this section, with a standard deviation of ${\sim}$0.01~K on the thermometer readings. The only parameter that is varied is the electron beam current density. 

Figure~\ref{fig:beam}(a) shows the resistance plotted as a function of time for an experiment with an average thermometer temperature of 6.1~K. The electron beam was blanked during the first 15~s, after which the beam was unblanked and the current density was decreased stepwise every 15~s by increasing the spot size. Imaging was conducted at low magnification so that the entire {$\sim$}0.5~mm diameter aperture was illuminated with the electron beam. A TEM image of the sample is shown in the inset to Fig.~3(b). Initially, at $t$=0-15s, the device is superconducting when the electron beam is blanked. When the beam is unblanked at $t$=15~s (current density $2.6 \times 10^{-2}$~\textit{e}\,s$^{-1}$\AA$^{-2}$), the resistance increases to ${\sim}0.4~\Omega$, corresponding to the device resistance above $T_{\mathrm{c}}$. The resistance subsequently decreases with decreasing beam current density, but does not become zero until the beam current density reaches $6 \times 10^{-4}$~\textit{e}\,s$^{-1}$\AA$^{-2}$. The device briefly becomes superconducting each time the spot size is changed because the beam is blanked for a few seconds. 

Figure~\ref{fig:beam}(b) shows the average device resistance plotted against electron beam current density for four different experiments performed at thermometer temperatures in the range 5.9–6.6~K. During each of these measurements, the temperature remains approximately constant, with a standard deviation of {$\sim$}0.01~K on the thermometer readings. The resistance generally increases with increasing electron beam current density. For all experiments, we find that the device reaches $\sim$0.4~$\Omega$, corresponding to the device resistance above $T_{\mathrm{c}}$, at the largest tested current density of $2.6 \times 10^{-2}$~\textit{e}\,s$^{-1}$\AA$^{-2}$. In the experiment with the highest thermometer temperature ($T$=6.6~K), the device reaches this normal-state resistance already at a beam current density of 1.1$\times 10^{-2}$~\textit{e}\,s$^{-1}$\AA$^{-2}$. 

\begin{figure*}[t]
	\includegraphics[width=1\linewidth]{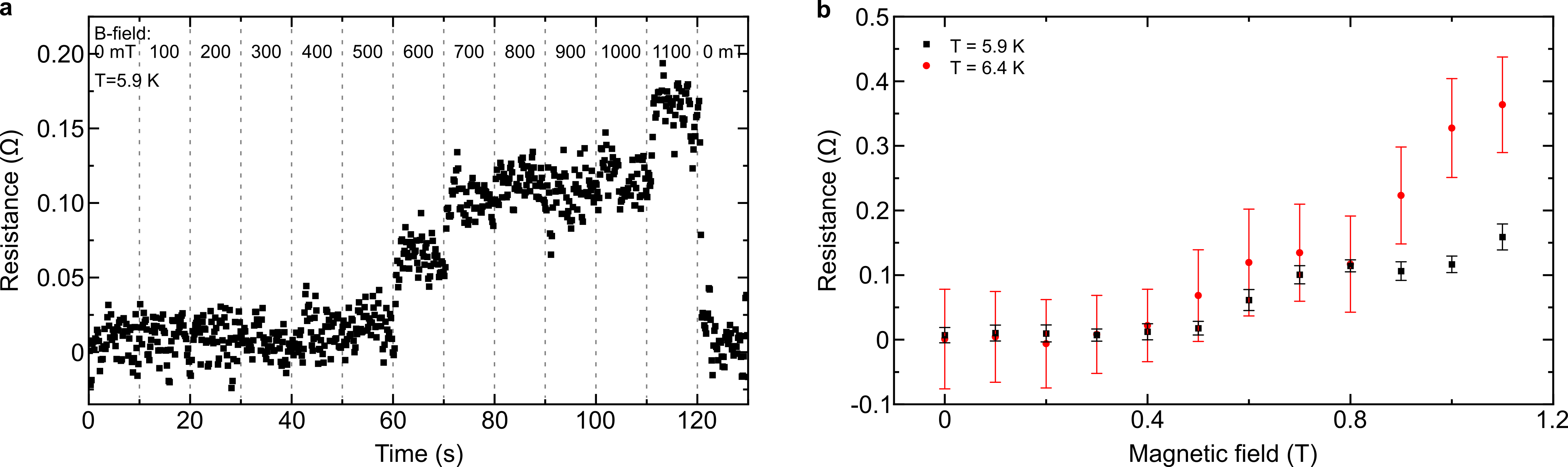}
	\caption{\label{fig:Bfield}
		\textbf{Effect of magnetic field.} 
		\textbf{(a)} Resistance plotted against time during an experiment where the electron beam is blanked, and the magnetic field is increased in steps of 100~mT every 10~s as indicated on the plot. The thermometer temperature was 5.9~K during the measurement. 
        \textbf{(b)} Average resistance plotted against magnetic field for two different experiments performed at thermometer readings of $T$=5.9 and 6.4~K. The error bars (the standard deviation of the measurements) are larger for the red data compared to the black data because test currents of 100 and 500~nA were used for the red and black data, respectively. 
  		}
\end{figure*}

Repeating the experiment at approximately 200~K, well above $T_{\mathrm{c}}$, did not result in any change in the device resistance between having the electron beam blanked and illuminating the device with a current density of $2.6 \times 10^{-2}$~\textit{e}\,s$^{-1}$\AA$^{-2}$, as shown in Fig.~S7, suggesting that the resistance changes are not due to changes in, \textit{e.g.}, electron-beam-induced doping or charging effects. 

We interpret the data as follows. The device exhibits an apparent $T_{\mathrm{c}}^{\ast}$=7.3~K (device 1, modified shield), whereas independent PPMS measurements give $T_{\mathrm{c}}=11.2$~K, implying that the actual specimen temperature is approximately 4~K higher than the thermometer reading. Thus, for thermometer readings between 5.9 and 6.6~K, the specimen temperature with the beam blanked is estimated to be $\sim$1~K below $T_{\mathrm{c}}$. For electron beam current densities below $6 \times 10^{-4}$~\textit{e}\,s$^{-1}$\AA$^{-2}$, we do not observe a measurable increase in resistance, indicating that electron-beam-induced heating is negligible and the device remains superconducting. At higher current densities, the specimen temperature is raised towards or above $T_{\mathrm{c}}$, depending on the TEM holder temperature during the experiment and the electron beam current density. The resistance, therefore, increases from zero towards the normal-state value of {$\sim$}0.4~$\Omega$ with increasing current density. This observation suggests a modest temperature rise of {$\sim$}1~K for the range of electron beam current densities tested. 

Specimen heating depends on specimen thickness, beam current, beam diameter, the thermal conductivity of the constituent materials, and the distance to available heat sinks \cite{egerton2004radiation}. In our case, the device is placed on a silicon nitride membrane, which has a thermal conductivity of less than 1~Wm$^{-1}$K$^{-1}$ at temperatures below 50~K \cite{ftouni2015thermal}. This is very low when compared to copper, whose thermal conductivity can exceed several thousand Wm$^{-1}$K$^{-1}$ at low temperatures \cite{ho1972thermal}, suggesting that electron-beam-induced heating may be smaller for samples mounted on conventional Cu TEM grids. However, electrical devices typically require silicon/silicon nitride support chips, on which electrodes and contact pads can be patterned. In order to reduce electron-beam-induced heating on such platforms, it may be necessary to introduce additional heat conduction pathways, for example \textit{via} nearby Cu heat-spreading lines or buried high-conductivity layers beneath the silicon nitride. Another factor is thermal resistance at the interfaces between the TEM grid and the cartridge, and the cartridge and the TEM sample holder. These contacts could be optimized to increase the thermal conductivity across the interfaces, for instance by using thermally conducting varnish. 

Regarding the beam current, we note that sample heating may be more severe under high-resolution TEM conditions, which can require local electron doses on the order of at least ${\sim}10^5$~\textit{e}/\AA$^{2}$, highlighting the importance of efficient heat transfer away from the specimen.

Regarding the specimen thickness, we note that the present device is completely opaque to the electron beam (see Fig.~3(b), inset--it is approximately 1~µm thick), representing a worst case scenario, in which the device absorbs most of the electron beam energy and can be heated significantly \cite{egerton2004radiation}. Measurements of electron transparent samples of Cu$_2$Se with a relative low thermal conductivity (0.2–0.9~Wm$^{-1}$K$^{-1}$) have revealed negligible heating at dose rates of a few hundred \textit{e}\,s$^{-1}$\AA$^{-2}$ \cite{chen2018temperature}. We anticipate that, all other factors being equal, heating effects will be reduced for electron transparent samples. For instance, using van der Waals heterostructures based on two-dimensional materials, it is possible to make ultra-thin devices, enabling atomic resolution imaging and simultaneous electrostatic gating or electric field control \cite{han2025electric, thomsen2024direct}. 

Finally, we evaluate the low-temperature imaging performance and mechanical stability of the sample holder. The NbN devices studied here are too thick for TEM imaging, and we also find that their superconducting properties are degraded by FIB processing. We therefore evaluate the imaging performance and stability of the holder from a time series of LTEM images of a skyrmion lattice in CrBr$_3$, a van der Waals ferromagnetic with a Curie temperature of $\sim$37~K \cite{grebenchuk2024topological}; further details on the drift analysis and sample are provided in Fig.~S8 and its caption. The time series is shown in Supplementary Video 1; details about the video are provided in Supplementary Note 2. In summary, the video shows that the specimen motion consists of both linear drift and superimposed vibrations. From the drift analysis, we extract linear drift rates of 3.7~nm/s and 1.2~nm/s in the \textit{x}- and \textit{y}-directions, respectively. The standard deviations of the vibrational motion are 3.2~nm and 1.8~nm in the \textit{x}- and \textit{y}-directions, respectively. Although this stability is not yet sufficient for atomic-resolution imaging, it is adequate for imaging skyrmions and magnetic domain walls in CrBr$_3$. This is in good agreement with a recent report on a continuous-flow liquid-helium-cooled holder from the same manufacturer, designed for a different TEM platform \cite{kim2025ultralow}. We anticipate that further developments can improve the mechanical stability of the sample holder. 

    \subsection{Objective lens excitation}

We now investigate the effect of an applied magnetic field generated by partial excitation of the objective lens. In these experiments we measure the resistance of NbN device~1 while using the modified cryo-shield in Lorentz TEM mode without electron beam illumination. The temperature is held constant during each experiment, with a standard deviation of ${\sim}$0.01~K on the thermometer readings. Only the magnetic field is varied through partial excitation of the objective lenses. 

Figure~\ref{fig:Bfield}(a) shows the resistance of NbN device~1 plotted as a function of time during an experiment with a thermometer reading of 6.1~K, with the magnetic field increased in steps of 100~mT every 10~s. The device remains superconducting up to a magnetic field of {$\sim$}400~mT. Above 400~mT, the resistance increases in a non-monotonic manner. After the magnetic field is removed at $t > 120$~s, the device returns to the superconducting state in less than 0.5~s.

Figure~\ref{fig:Bfield}(b) shows the average resistance plotted against magnetic field for the data shown in Fig.~\ref{fig:Bfield}(a) and another experiment performed at $T$=6.4~K. 
For both experiments we find that the resistance increases above 400 mT. In addition, the resistance above 400 mT is higher for the data obtained at 6.4~K compared to 5.9~K. 

In the following, we discuss several possible origins of the observed increase in device resistance with increasing magnetic field. One possibility is a temperature rise in the objective lens during excitation, which could increase thermal radiation onto the specimen and thereby raise the device temperature. Another possibility is related to the critical magnetic field of NbN. Near 0~K, the critical field exceeds 10~T, but it decreases and eventually approaches zero as the temperature approaches \(T_{\mathrm{c}}\) \cite{hwang2015transition}. The observed increase in resistance with applied magnetic field can therefore be understood if the device temperature is close to \(T_{\mathrm{c}}\), such that partial objective-lens excitation is sufficient to exceed the critical field. The higher resistance of the red dataset compared with the black dataset in Fig.~\ref{fig:Bfield}(b) for fields above 400~mT may then reflect a lower critical field at the slightly higher temperature. Finally, flux-flow dissipation of superconducting vortices can also produce a finite resistance under applied magnetic field \cite{kim1965flux}. 

We note that NbN device 2 exhibits a critical field of 2.8~T at 10 K (Fig.~S6), which exceeds the tested magnetic field strengths in Fig.~\ref{fig:Bfield}. However, the TEM measurement was performed on NbN device 1 at temperatures around 6 K. Since the two devices differ in thickness and \(T_{\mathrm{c}}\), their critical magnetic fields may also differ. Nevertheless, this comparison indicates that objective-lens heating may be an important contributor to the apparent field-dependent resistance increase, although a reduced critical field in device 1 and flux-flow dissipation cannot be ruled out.

\subsection{Thermal radiation calculations}

In this section, we estimate the radiative heat loads for different cryo-shield configurations to clarify the role of cryo-shielding. We distinguish between three contributions: (1) direct line-of-sight radiative exchange between the device and the room-temperature surroundings through the imaging apertures; (2) room-temperature radiation admitted through the apertures that is absorbed by the inner surfaces of the cryo-shield and the TEM grid and can subsequently warm the device \textit{via} thermal diffusion; and (3) radiation absorbed by the outside of the cryo-shield from surrounding warm surfaces. Contributions (1) and (2) depend strongly on imaging aperture size, whereas (3) is primarily set by the shield’s external view and surface properties. Our analysis suggests that the indirect pathways (2) and (3) can dominate the observed device heating, depending on the thermal coupling between the device, TEM grid, and cryo-shield.

Starting with the line-of-sight contribution, we estimate the direct net radiative heat load using the Stefan--Boltzmann law,

\begin{equation}
P = \epsilon \sigma 2A(T^{4}-T_{\mathrm{d}}^{4}),
\end{equation}

where $\epsilon$ is the effective emissivity, $\sigma$ the Stefan-Boltzmann constant, $A$ the device surface area of one face, $T$ the temperature of the radiating surroundings, and $T_{\mathrm{d}}$ the device temperature. The factor of 2 accounts for radiative exchange on both the top and the bottom face of the device. We furthermore assume that $\epsilon=1$ (blackbody limit) and $T_{\mathrm{d}}$=10 K. 

We first consider the case without a cryo-shield. The device (area $70\times150~\mu$m$^2$) has an unobstructed view of room-temperature surfaces on both sides. Under these conditions, the direct net radiative heat load is $P_{\mathrm{dir,noshield}} \sim 10~\mu$W. 

Next, we consider the effect of the imaging aperture in the cryo-shield. Viewing apertures can be a major contributor to radiative heat load in cryostats \cite{jacques2025cryogenic}. This situation requires consideration of the view factor, which quantifies the fraction of radiation leaving one surface that reaches another. For diffuse surfaces, the view factor from surface 1 (area A$_1$) to surface 2 (area A$_2$) is \cite{incropera2013principles}

\begin{figure}[t]
	\includegraphics[width=1\linewidth]{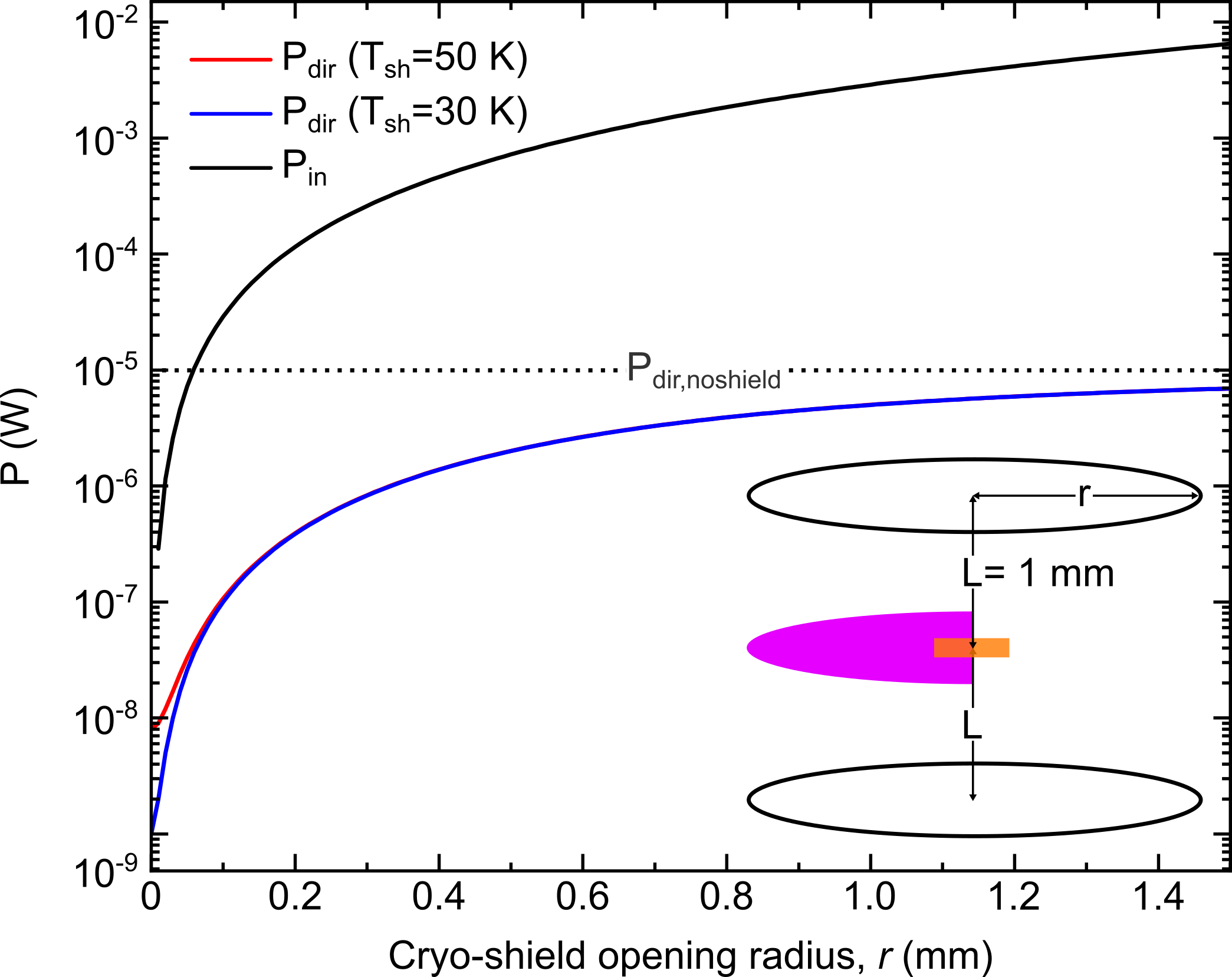}
	\caption{\label{fig:calc}
		\textbf{Thermal radiation heat loads.} 
		      The red and blue curves show the net radiative power (direct line-of-sight) on the device, $P_{\mathrm{dir}}$, calculated using Eq.~4, for $T_{\mathrm{sh}}=30$ K (blue) and 50 K (red). The radiative power transmitted through the imaging apertures,  $P_{\mathrm{in}}$, is plotted in black. The inset shows a diagram of the model. The horizontal dotted line indicates the direct line-of-sight contribution without a cryo-shield, $P_{\mathrm{dir,noshield}}$. 
  		}
\end{figure}

\begin{equation}
F_{1\rightarrow 2} = \frac{1}{A_1}\int_{A_1}\int_{A_2}
\frac{\cos\theta_1\,\cos\theta_2}{\pi s^{2}}\,\mathrm{d}A_{2}\,\mathrm{d}A_{1},
\end{equation}

where $\theta_1$ and $\theta_2$ are the angles between the surface normals and the line connecting the differential areas, and $s$ is the distance between the two differential areas. Approximating the device as a point-like surface facing a parallel circular aperture of radius $r$ at distance $L$ (see Fig.~5, inset), the view factor reduces to \cite{naraghi1988radiation}

\begin{equation}
    F_{1\rightarrow 2} = \frac{r^2}{L^{2}+r^{2}}.
\end{equation}

The derivation is provided in Supplementary Note 1. We assume identical coaxial openings above and below the device, such that the same $F_{1\rightarrow 2}$ applies to both faces. The total net radiative power absorbed by the device (direct line-of-sight contribution) is then

\begin{equation}
    P_{\mathrm{dir}} = P_{\mathrm{dir,noshield}}\,F_{1\rightarrow 2} + P_{\mathrm{sh}}(1-F_{1\rightarrow 2}),
\end{equation}

where $P_{\mathrm{sh}}$ is the net radiative exchange with the shield at temperature $T_{\mathrm{sh}}$, evaluated using Eq.~1 with $T = T_{\mathrm{sh}}$. Figure~5 shows a plot of $P_{\mathrm{dir}}$ as a function of the radius of the cryo-shield openings, assuming $T_{\mathrm{sh}}$=30~K and 50~K. This treatment assumes diffuse radiative exchange, with the opening providing direct view to room-temperature surroundings and the remaining solid angle occupied by the shield interior.

We find that, for any practical cryo-shield opening radius, the assumed $T_{\mathrm{sh}}$ has only a minor effect on the net radiative power absorbed by the device. For $r=0$, corresponding to a fully closed imaging aperture, the device views only the cryo-shield and the net radiative power is $\sim$1~nW and $\sim$8~nW for $T_{\mathrm{sh}}$=30~K and 50~K, respectively. Hence, a fully closed imaging aperture reduces the direct line-of-sight radiative load on the device by 3–4 orders of magnitude when compared to not having a cryo-shield, depending on $T_{\mathrm{sh}}$.

The radiative power admitted through the imaging apertures (contribution (2)) can be estimated in the blackbody limit by applying the Stefan–Boltzmann law to the area of the two imaging apertures,
\begin{equation}
P_{\mathrm{in}} \approx A_{\mathrm{holes}}\sigma\left(T_{\mathrm{room}}^{4}-T_{\mathrm{sh}}^{4}\right),\qquad A_{\mathrm{holes}}=2\pi r^{2},
\end{equation}
which scales as $P_{\mathrm{in}}\propto r^{2}$. For $r=0.25$~mm and 1.5~mm, this calculation yields $P_{\mathrm{in}}{\sim}$0.2~mW and {$\sim$}6.5~mW, respectively. $P_{\mathrm{in}}$ is plotted as a function of $r$ in Fig.~5. The fraction of $P_{\mathrm{in}}$ that contributes to heating depends on how much is absorbed by the grid/holder/shield surfaces and on internal reflections within the cryo-shield.

Finally, thermal radiation absorbed by the outside of the cryo-shield (contribution (3)) can provide an additional heat load. Because the external surface area of the cryo-shield is larger than the aperture area, this contribution can be comparable to or larger than $P_{\mathrm{in}}$. However, its magnitude depends on the outside emissivity and the view factor of the shield exterior to room-temperature surroundings. Moreover, because this heat is deposited further from the device, its impact on the device temperature can be reduced by weaker thermal coupling to the device.

We note that both $P_{\mathrm{dir}}$ and $P_{\mathrm{in}}$ decrease with decreasing $r$. The calculation of $P_{\mathrm{dir}}$ quantifies the maximum radiative power that can be absorbed directly by the device through line-of-sight to room-temperature surfaces, providing an upper bound on this pathway. In contrast, $P_{\mathrm{in}}$ represents a radiative power budget entering the cryo-shield volume that may be absorbed by the cryo-shield, TEM grid, or sample holder and subsequently transferred to the device by conduction. Only a small fraction of $P_{\mathrm{in}}$ needs to be absorbed and thermally coupled to the device for indirect heating to dominate, since $P_{\mathrm{dir}}$/$P_{\mathrm{in}}\sim 0.6$ µW/0.2 mW = $3\times10^{-3}$ for $r=0.25$ mm.

Overall, these estimates show that the imaging aperture is an important factor to consider in such experiments, in agreement with our experimental observations. The as-supplied cryo-shield with $r{=}1.5$~mm has $F_{1\rightarrow 2}{\sim}$0.7, and provides only $\sim$30\% reduction of the direct radiative load from the room-temperature surroundings. For the modified cryo-shield with $r{=}0.25$~mm, the view factor is $F_{1\rightarrow 2}{\sim}$0.06, corresponding to $\sim$94\% reduction. The aperture-admitted radiative power is likewise reduced by 97\% for the modified cryo-shield. 

\section{Conclusions and outlook}

We have demonstrated electrical transport measurements on a superconducting NbN device inside a TEM using a continuous-flow liquid-helium sample holder. The device exhibits an apparent $T_{\mathrm{c}}$ that depends sensitively on the cryo-shielding configuration and on sources of thermal radiation. By using a modified cryo-shield with an aperture of $\sim$0.5~mm for the electron beam, we obtain an apparent $T_{\mathrm{c}}$ of 7.3~K. When combined with the intrinsic $T_{\mathrm{c}}=11.2$~K and a base thermometer reading of 4.5~K, this observation suggests that the lowest specimen temperature attainable is approximately 8–9~K. In addition, we find that electron beam illumination leads to specimen heating, pointing to the importance of efficient heat conductance away from the specimen. Finally, excitation of the objective lens can drive the device out of the superconducting state, possibly due to increased thermal radiation caused by lens heating, although direct magnetic-field effects and flux-flow dissipation cannot be ruled out.

Our results demonstrate that, during TEM imaging without efficient heat shielding, the specimen temperature is \textit{at minimum} 5~K higher than the thermometer reading in the present specimen sample holder. This finding indicates that precise specimen temperature calibration is essential for accurately quantifying the functional properties of quantum materials during liquid-helium-cooled TEM measurements. Such calibration may be enabled by novel MEMS-style heater chips that can operate at liquid-helium temperatures. Further reductions in specimen temperature will require optimizing both thermal radiation shielding and heat transport away from the specimen. Calculations show that it is essential to minimize the imaging apertures to minimize thermal radiation admitted into the sample space. 

Future studies should investigate the use of cryo-blades or cryo-cages around the specimen, as well as optimized TEM grids incorporating high-conductivity heat-spreading pathways, since the silicon-nitride-based membranes used here provide poor thermal transport. Experiments on electron-transparent samples with $T_{\mathrm{c}}$ close to the base temperature of the holder will be needed to evaluate beam effects more precisely, particularly at the higher electron doses required for high-resolution TEM. One possible route would be to develop electron-transparent NbN films ($\sim$100~nm thick), either through an optimized FIB thinning process or by direct growth on an electron-transparent crystalline substrate such as a thin silicon carbide membrane. Alternatively, flux vortices have been observed by Lorentz TEM in Nb thin films prepared by chemical etching of rolled foils, which have $T_{\mathrm{c}}$=9.2~K \cite{harada1996direct}. If the base temperature of the sample holder can be reduced by $\sim$2 K, NbSe$_2$ would also be an attractive system, since it retains $T_{\mathrm{c}}$=7.2~K down to a thickness of only a few nanometers and offers the additional opportunity to study the interplay between charge-density-wave and superconducting order in exfoliated single-crystalline flakes \cite{xi2015strongly}.

These results demonstrate the feasibility of combining superconducting transport measurements with TEM observation. This capability opens the door to future experiments that directly correlate microstructural, spectroscopic, and electronic properties in superconductors and quantum devices, enabling a deeper understanding of how defects, strain, and interfaces influence superconductivity at the nanoscale. Such observations could play a key role in guiding the design of next-generation quantum materials and devices.

\section*{Declaration of competing interest}
D.B. and D.S. are involved in developing cryogenic electron microscopy sample holders at condenZero AG. The remaining authors declare no competing financial interests or personal relationships that could have appeared to influence the work reported in this paper.

\section*{Acknowledgements}
This work was supported by the European Union’s Horizon 2020 Research and Innovation Programme (grant 856538, project 3D MAGIC). R.B.D.\ acknowledges support from the European Union under grant agreement no. 101094299 (project IMPRESS). Part of the work was performed at JCNS-2 and PGI-7 in Forschungszentrum Jülich. The authors thank A.\ Kovács, R.\ Borowski, L.\ Kibkalo, L.\ Risters, O.\ Petracic, S.\ Nandi, and B.\ Schmitz for technical assistance. The authors would like to acknowledge C.\ Stampfer and B.\ Beschoten (RWTH Aachen) for access to their custom-built van der Waals transfer system for
heterostructure assembly.
 
\section*{Data availability}
Data will be made available on reasonable request.

\bibliography{references}

@article{tokura2017emergent,
  title={Emergent functions of quantum materials},
  author={Tokura, Yoshinori and Kawasaki, Masashi and Nagaosa, Naoto},
  journal={Nature Physics},
  volume={13},
  number={11},
  pages={1056--1068},
  year={2017},
  publisher={Nature Publishing Group UK London}
}

@article{moler2017imaging,
  title={Imaging quantum materials},
  author={Moler, Kathryn Ann},
  journal={Nature Materials},
  volume={16},
  number={11},
  pages={1049--1052},
  year={2017},
  publisher={Nature Publishing Group UK London}
}

@article{minor2019cryogenic,
  title={Cryogenic electron microscopy for quantum science},
  author={Minor, Andrew M and Denes, Peter and Muller, David A},
  journal={MRS Bulletin},
  volume={44},
  number={12},
  pages={961--966},
  year={2019},
  publisher={Cambridge University Press}
}

@article{bianco2021atomic,
  title={Atomic-resolution cryogenic scanning transmission electron microscopy for quantum materials},
  author={Bianco, Elisabeth and Kourkoutis, Lena F},
  journal={Accounts of Chemical Research},
  volume={54},
  number={17},
  pages={3277--3287},
  year={2021},
  publisher={ACS Publications}
}

@article{hart2021seeing,
  title={Seeing quantum materials with cryogenic transmission electron microscopy},
  author={Hart, James L and Cha, Judy J},
  journal={Nano Letters},
  volume={21},
  number={13},
  pages={5449--5452},
  year={2021},
  publisher={ACS Publications}
}

@article{han2019topological,
  title={Topological magnetic-spin textures in two-dimensional van der {W}aals {Cr$_2$Ge$_2$Te$_6$}},
  author={Han, Myung-Geun and Garlow, Joseph A and Liu, Yu and Zhang, Huiqin and Li, Jun and DiMarzio, Donald and Knight, Mark W and Petrovic, Cedomir and Jariwala, Deep and Zhu, Yimei},
  journal={Nano Letters},
  volume={19},
  number={11},
  pages={7859--7865},
  year={2019},
  publisher={ACS Publications}
}

@article{xi2015strongly,
  title={Strongly enhanced charge-density-wave order in monolayer {NbSe$_2$}},
  author={Xi, Xiaoxiang and Zhao, Liang and Wang, Zefang and Berger, Helmuth and Forr{\'o}, L{\'a}szl{\'o} and Shan, Jie and Mak, Kin Fai},
  journal={Nature Nanotechnology},
  volume={10},
  number={9},
  pages={765--769},
  year={2015},
  publisher={Nature Publishing Group UK London}
}

@article{chen2022lorentz,
  title={Lorentz electron ptychography for imaging magnetic textures beyond the diffraction limit},
  author={Chen, Zhen and Turgut, Emrah and Jiang, Yi and Nguyen, Kayla X and Stolt, Matthew J and Jin, Song and Ralph, Daniel C and Fuchs, Gregory D and Muller, David A},
  journal={Nature Nanotechnology},
  volume={17},
  number={11},
  pages={1165--1170},
  year={2022},
  publisher={Nature Publishing Group UK London}
}

@article{cantoni2014orbital,
  title={Orbital Occupancy and Charge Doping in Iron-Based Superconductors},
  author={Cantoni, Claudia and Mitchell, Jonathan E and May, Andrew F and McGuire, Michael A and Idrobo, Juan-Carlos and Berlijn, Tom and Dagotto, Elbio and Chisholm, Matthew F and Zhou, Wu and Pennycook, Stephen J and others},
  journal={Advanced Materials},
  volume={26},
  number={35},
  pages={6193--6198},
  year={2014},
  publisher={Wiley Online Library}
}

@article{zhao2018direct,
  title={Direct imaging of electron transfer and its influence on superconducting pairing at {FeSe/SrTiO$_3$} interface},
  author={Zhao, Weiwei and Li, Mingda and Chang, Cui-Zu and Jiang, Jue and Wu, Lijun and Liu, Chaoxing and Moodera, Jagadeesh S and Zhu, Yimei and Chan, Moses HW},
  journal={Science Advances},
  volume={4},
  number={3},
  pages={eaao2682},
  year={2018},
  publisher={American Association for the Advancement of Science}
}

@article{harada1992real,
  title={Real-time observation of vortex lattices in a superconductor by electron microscopy},
  author={Harada, K and Matsuda, T and Bonevich, J and Igarashi, M and Kondo, S and Pozzi, G and Kawabe, U and Tonomura, A},
  journal={Nature},
  volume={360},
  number={6399},
  pages={51--53},
  year={1992},
  publisher={Nature Publishing Group UK London}
}

@article{schneider2010wedging,
  title={Wedging transfer of nanostructures},
  author={Schneider, Gr{\'e}gory F and Calado, Victor E and Zandbergen, Henny and Vandersypen, Lieven MK and Dekker, Cees},
  journal={Nano Letters},
  volume={10},
  number={5},
  pages={1912--1916},
  year={2010},
  publisher={ACS Publications}
}

@article{ali2025visualizing,
  title={Visualizing subatomic orbital and spin moments using a scanning transmission electron microscope},
  author={Ali, Hasan and Rusz, Jan and B{\"u}rgler, Daniel E and Vas, Joseph V and Jin, Lei and Adam, Roman and Schneider, Claus M and Dunin-Borkowski, Rafal E},
  journal={Nature Materials},
  pages={1--6},
  year={2025},
  publisher={Nature Publishing Group UK London}
}

@article{hwang2015transition,
  title={Transition temperatures and upper critical fields of {N}b{N} thin films fabricated at room temperature},
  author={Hwang, TJ and Kim, DH},
  journal={Progress in Superconductivity and Cryogenics},
  volume={17},
  number={3},
  pages={9--12},
  year={2015},
  publisher={The Korean Society of Superconductivity and Cryogenics}
}

@article{guo2020fabrication,
  title={Fabrication of superconducting niobium nitride nanowire with high aspect ratio for {X}-ray photon detection},
  author={Guo, Shuya and Chen, Qi and Pan, Danfeng and Wu, Yaojun and Tu, Xuecou and He, Guanglong and Han, Hang and Li, Feiyan and Jia, Xiaoqing and Zhao, Qingyuan and others},
  journal={Scientific Reports},
  volume={10},
  number={1},
  pages={9057},
  year={2020},
  publisher={Nature Publishing Group UK London}
}

@article{hazra2016superconducting,
  title={Superconducting properties of very high quality {N}b{N} thin films grown by high temperature chemical vapor deposition},
  author={Hazra, D and Tsavdaris, N and Jebari, S and Grimm, Alexander and Blanchet, F and Mercier, Florian and Blanquet, E and Chapelier, C and Hofheinz, M},
  journal={Superconductor Science and Technology},
  volume={29},
  number={10},
  pages={105011},
  year={2016},
  publisher={IOP Publishing}
}

@article{volkov2019superconducting,
  title={Superconducting properties of very high quality {N}b{N} thin films grown by pulsed laser deposition},
  author={Volkov, Serhii and Gregor, Maros and Roch, Tomas and Satrapinskyy, Leonid and Gran{\v{c}}i{\v{c}}, Branislav and Fiantok, Tomas and Plecenik, Andrej},
  journal={Journal of Electrical Engineering},
  volume={70},
  number={7S},
  pages={89--94},
  year={2019},
  publisher={De Gruyter Poland}
}

@article{mathur1972lower,
  title={Lower critical field measurements in {N}b{N} bulk and thin films},
  author={Mathur, MP and Deis, DW and Gavaler, JR},
  journal={Journal of Applied Physics},
  volume={43},
  number={7},
  pages={3158--3161},
  year={1972},
  publisher={American Institute of Physics}
}

@article{kim2025ultralow,
  title={Ultralow-temperature cryogenic transmission electron microscopy using a new helium flow cryostat stage},
  author={Kim, Young-Hoon and Yasin, Fehmi Sami and Kim, Na Yeon and Birch, Max and Yu, Xiuzhen and Kikkawa, Akiko and Taguchi, Yasujiro and Yan, Jiaqiang and Chi, Miaofang},
  journal={Ultramicroscopy},
  pages={114263},
  year={2025},
  publisher={Elsevier}
}

@article{kang2025large,
  title={Large-angle {L}orentz Four-dimensional scanning transmission electron microscopy for simultaneous local magnetization, strain and structure mapping},
  author={Kang, Sangjun and T{\"o}llner, Maximilian and Wang, Di and Minnert, Christian and Durst, Karsten and Caron, Arnaud and Dunin-Borkowski, Rafal E and McCord, Jeffrey and K{\"u}bel, Christian and Mu, Xiaoke},
  journal={Nature Communications},
  volume={16},
  number={1},
  pages={1305},
  year={2025},
  publisher={Nature Publishing Group UK London}
}

@article{han2025electric,
  title={Electric Field Control of Magnetic Skyrmion Helicity in a Centrosymmetric 2{D} van der {W}aals Magnet},
  author={Han, Myung-Geun and Thomsen, Joachim Dahl and Philbin, John P and Mun, Junsik and Park, Eugene and Camino, Fernando and Dekanovsky, Lukas and Liu, Chuhang and Sofer, Zdenek and Narang, Prineha and others},
  journal={Nano Letters},
  volume={25},
  number={13},
  pages={5174--5180},
  year={2025},
  publisher={ACS Publications}
}

@article{midgley2009electron,
  title={Electron tomography and holography in materials science},
  author={Midgley, Paul A and Dunin-Borkowski, Rafal E},
  journal={Nature Materials},
  volume={8},
  number={4},
  pages={271--280},
  year={2009},
  publisher={Nature Publishing Group UK London}
}

@article{ko2023understanding,
  title={Understanding heterogeneities in quantum materials},
  author={Ko, Wonhee and Gai, Zheng and Puretzky, Alexander A and Liang, Liangbo and Berlijn, Tom and Hachtel, Jordan A and Xiao, Kai and Ganesh, Panchapakesan and Yoon, Mina and Li, An-Ping},
  journal={Advanced Materials},
  volume={35},
  number={27},
  pages={2106909},
  year={2023},
  publisher={Wiley Online Library}
}

@article{mazza2024embracing,
  title={Embracing disorder in quantum materials design},
  author={Mazza, Alessandro R and Yan, J-Q and Middey, S and Gardner, JS and Chen, A-H and Brahlek, M and Ward, T Zac},
  journal={Applied Physics Letters},
  volume={124},
  number={23},
  year={2024},
  publisher={AIP Publishing}
}

@article{mun2024atomic,
  title={Atomic resolution scanning transmission electron microscopy at liquid helium temperatures for quantum materials},
  author={Mun, Junsik and Potemkin, Daniel and Jang, Houk and Park, Suji and Mick, Stephen and Petrovic, Cedomir and Cheong, Sang-Wook and Han, Myung-Geun and Zhu, Yimei},
  journal={Ultramicroscopy},
  volume={267},
  pages={114039},
  year={2024},
  publisher={Elsevier}
}

@article{rennich2025ultracold,
  title={Ultracold cryogenic {TEM} with liquid helium and high stability},
  author={Rennich, Emily and Sung, Suk Hyun and Agarwal, Nishkarsh and Gates, Maya and Kerns, Robert and Hovden, Robert and Baggari, Ismail El},
  journal={Proceedings of the National Academy of Sciences},
  volume={122},
  number={36},
  pages={e2509736122},
  year={2025},
  publisher={National Academy of Sciences}
}

@article{kumar2024calibrating,
  title={Calibrating cryogenic temperature of {TEM} specimens using {EELS}},
  author={Kumar, Abinash and Tiukalova, Elizaveta and Venkatraman, Kartik and Lupini, Andrew and Hachtel, Jordan A and Chi, Miaofang},
  journal={Ultramicroscopy},
  volume={265},
  pages={114008},
  year={2024},
  publisher={Elsevier}
}

@article{lefranc1982superconducting,
  title={Superconducting lens design},
  author={Lefranc, G and Knapek, E and Dietrich, I},
  journal={Ultramicroscopy},
  volume={10},
  number={1-2},
  pages={111--123},
  year={1982},
  publisher={Elsevier}
}

@article{weierstall1996electron,
  title={Electron holography with a superconducting objective lens},
  author={Weierstall, Uwe and Lichte, H},
  journal={Ultramicroscopy},
  volume={65},
  number={1-2},
  pages={13--22},
  year={1996},
  publisher={Elsevier}
}

@article{egerton2004radiation,
  title={Radiation damage in the {TEM} and {SEM}},
  author={Egerton, Ray F and Li, Peng and Malac, Marek},
  journal={Micron},
  volume={35},
  number={6},
  pages={399--409},
  year={2004},
  publisher={Elsevier}
}

@article{ftouni2015thermal,
  title={Thermal conductivity of silicon nitride membranes is not sensitive to stress},
  author={Ftouni, Hossein and Blanc, Christophe and Tainoff, Dimitri and Fefferman, Andrew D and Defoort, Martial and Lulla, Kunal J and Richard, Jacques and Collin, Eddy and Bourgeois, Olivier},
  journal={Physical Review B},
  volume={92},
  number={12},
  pages={125439},
  year={2015},
  publisher={APS}
}

@article{ho1972thermal,
  title={Thermal conductivity of the elements},
  author={Ho, Cho Yen and Powell, Reginald W and Liley, Peter E},
  journal={Journal of Physical and Chemical Reference Data},
  volume={1},
  number={2},
  pages={279--421},
  year={1972},
  publisher={American Institute of Physics for the National Institute of Standards and~…}
}

@article{goodge2020atomic,
  title={Atomic-resolution cryo-{STEM} across continuously variable temperatures},
  author={Goodge, Berit H and Bianco, Elisabeth and Schnitzer, Noah and Zandbergen, Henny W and Kourkoutis, Lena F},
  journal={Microscopy and Microanalysis},
  volume={26},
  number={3},
  pages={439--446},
  year={2020},
  publisher={Oxford University Press}
}

@article{hart2023operando,
  title={In operando cryo-{STEM} of pulse-induced charge density wave switching in {T}aS$_2$},
  author={Hart, James L and Siddique, Saif and Schnitzer, Noah and Funni, Stephen D and Kourkoutis, Lena F and Cha, Judy J},
  journal={Nature Communications},
  volume={14},
  number={1},
  pages={8202},
  year={2023},
  publisher={Nature Publishing Group UK London}
}

@article{li2013electron,
  title={Electron counting and beam-induced motion correction enable near-atomic-resolution single-particle cryo-{EM}},
  author={Li, Xueming and Mooney, Paul and Zheng, Shawn and Booth, Christopher R and Braunfeld, Michael B and Gubbens, Sander and Agard, David A and Cheng, Yifan},
  journal={Nature Methods},
  volume={10},
  number={6},
  pages={584--590},
  year={2013},
  publisher={Nature Publishing Group US New York}
}

@article{boothroyd2016fei,
  title={{FEI Titan G2 60-300 HOLO}},
  author={Boothroyd, Chris and Kov{\'a}cs, Andr{\'a}s and Tillmann, Karsten},
  journal={Journal of Large-scale Research Facilities},
  volume={2},
  pages={A44--A44},
  year={2016}
}

@article{incropera2013principles,
  title={Principles of heat and mass transfer},
  author={Incropera, Frank P and Dewitt, David P and Bergman, Theodore L and Lavine, Adrienne S},
  journal={(Wiley \& Sons Ltd)},
  year={2013}
}

@article{thomsen2024direct,
  title={Direct visualization of defect-controlled diffusion in van der {W}aals gaps},
  author={Thomsen, Joachim Dahl and Wang, Yaxian and Flyvbjerg, Henrik and Park, Eugene and Watanabe, Kenji and Taniguchi, Takashi and Narang, Prineha and Ross, Frances M},
  journal={Advanced Materials},
  volume={36},
  number={39},
  pages={2403989},
  year={2024},
  publisher={Wiley Online Library}
}

@article{chen2018temperature,
  title={Temperature distribution of wedge-shaped specimen in {TEM}},
  author={Chen, Lu and Wang, Yong and Zhang, Ze},
  journal={Micron},
  volume={110},
  pages={46--49},
  year={2018},
  publisher={Elsevier}
}

@article{jacques2025cryogenic,
  title={Cryogenic radiative cooling of a large payload for gravitational wave detector},
  author={Jacques, Lionel and Zeoli, Morgane and Amorosi, Anthony and Bertolini, Alessandro and Collette, Christophe and Cornelissen, Robin and Di Fronzo, Chiara and Habraken, Serge and Loicq, J and others},
  journal={Cryogenics},
  volume={147},
  pages={104057},
  year={2025},
  publisher={Elsevier}
}

@article{naraghi1988radiation,
  title={Radiation view factors from differential plane sources to disks-A general formulation},
  author={Naraghi, MHN},
  journal={Journal of thermophysics and heat transfer},
  volume={2},
  number={3},
  pages={271--274},
  year={1988}
}

@article{boffi1976cleavage,
  title={On the cleavage energy of magnesium oxide},
  author={Boffi, S and Ricci, M},
  journal={Materials Chemistry},
  volume={1},
  number={4},
  pages={289--296},
  year={1976},
  publisher={Elsevier}
}

@article{recalde2024weld,
  title={Weld-free mounting of lamellae for electrical biasing operando TEM},
  author={Recalde-Benitez, Oscar and Pivak, Yevheniy and Jiang, Tianshu and Winkler, Robert and Zintler, Alexander and Adabifiroozjaei, Esmaeil and Komissinskiy, Philipp and Alff, Lambert and Hubbard, William A and Perez-Garza, H Hugo and others},
  journal={Ultramicroscopy},
  volume={260},
  pages={113939},
  year={2024},
  publisher={Elsevier}
}

@article{grebenchuk2024topological,
  title={Topological spin textures in an insulating van der Waals ferromagnet},
  author={Grebenchuk, Sergey and McKeever, Conor and Grzeszczyk, Magdalena and Chen, Zhaolong and {\v{S}}i{\v{s}}kins, Makars and McCray, Arthur RC and Li, Yue and Petford-Long, Amanda K and Phatak, Charudatta M and Ruihuan, Duan and others},
  journal={Advanced Materials},
  volume={36},
  number={24},
  pages={2311949},
  year={2024},
  publisher={Wiley Online Library}
}

@article{harada1996direct,
  title={Direct observation of vortex dynamics in superconducting films with regular arrays of defects},
  author={Harada, K and Kamimura, O and Kasai, H and Matsuda, T and Tonomura, A and Moshchalkov, VV},
  journal={Science},
  volume={274},
  number={5290},
  pages={1167--1170},
  year={1996},
  publisher={American Association for the Advancement of Science}
}

@article{kim1965flux,
  title={Flux-flow resistance in type-II superconductors},
  author={Kim, YB and Hempstead, CF and Strnad, AR},
  journal={Physical Review},
  volume={139},
  number={4A},
  pages={A1163},
  year={1965},
  publisher={APS}
}

@article{sadki2004focused,
  title={Focused-ion-beam-induced deposition of superconducting nanowires},
  author={Sadki, ES and Ooi, Shuuichi and Hirata, Kazuto},
  journal={Applied Physics Letters},
  volume={85},
  number={25},
  pages={6206--6208},
  year={2004},
  publisher={AIP Publishing}
}

@article{li2008tunability,
  title={Tunability of the superconductivity of tungsten films grown by focused-ion-beam direct writing},
  author={Li, Wuxia and Fenton, JC and Wang, Yiqian and McComb, DW and Warburton, PA},
  journal={Journal of Applied Physics},
  volume={104},
  number={9},
  year={2008},
  publisher={AIP Publishing}
}
\bibliographystyle{naturemag}

\end{document}